\documentclass[preprint,showpacs,preprintnumbers,amsmath,amssymb]{revtex4-1}
\usepackage{mathrsfs}
\usepackage{graphicx}
\usepackage{dcolumn}
\usepackage{bm}

\begin{document}

\title{On the angular momentum of photons: effects of transversality condition on the quantization of radiation fields}

\author{Chun-Fang Li\footnote{Email address: cfli@shu.edu.cn}}

\affiliation{Department of Physics, Shanghai University, 99 Shangda Road, 200444
Shanghai, China}

\date{\today}

\begin{abstract}

The notion of intrinsic system of coordinates is introduced for the photon from the constraint of transversality condition. The degree of freedom to specify the intrinsic system is extracted from the same constraint, which turns out to be responsible for the spin Hall effect of light. It is shown that the fundamental quantization conditions that break down in the laboratory system of coordinates restore in the intrinsic system, which make it realizable to canonically quantize the radiation field. It is also shown that the dependence of the intrinsic system on the momentum underlies the noncommutativity of photon position in the laboratory system. The commutation relations of the spin and orbital angular momentum that were found by van Enk and Nienhuis [J. Mod. Opt. \textbf{41}, 963 (1994)] in a second quantized theory are re-derived in the present first quantized theory.

\end{abstract}

\pacs{42.50.Tx, 03.65.Ca, 42.90.+m}
\maketitle


\section{Introduction}

It is well known that the canonical commutation relations
\begin{subequations}\label{FCR}
\begin{align}
  [\hat{X}_i, \hat{X}_j] & = 0, \label{FCR-x's} \\
  [\hat{P}_i, \hat{P}_j] & = 0, \label{FCR-p's} \\
  [\hat{X}_i, \hat{P}_j] & = i \hbar \delta_{ij}, \label{FCR-xp}
\end{align}
\end{subequations}
between the position $\hat{\mathbf X}$ and the momentum $\hat{\mathbf P}$ form the cornerstone of quantum mechanics and are called by Dirac \cite{Dirac} the ``fundamental quantum conditions'', from which follows the canonical commutation relation of the orbital angular momentum (OAM) $\hat{\mathbf L} =\hat{\mathbf X} \times \hat{\mathbf P}$,
\begin{equation}\label{FCR-L's}
[\hat{L}_i, \hat{L}_j] =i \hbar \epsilon_{ijk} \hat{L}_k.
\end{equation}
Quantities $\hat{\mathbf X}$ and $\hat{\mathbf P}$ satisfying commutation relations (\ref{FCR}) are said to be canonically conjugate to each other. Quantities satisfying commutation relation (\ref{FCR-L's}) are rotation generators \cite{Sakurai}.
However, it was recognized \cite{Pryce, Skagerstam, Berard2006} that the Cartesian components of the photon position do not commute,
\begin{equation}\label{noncommute}
    [\hat{X}_i, \hat{X}_j] \neq 0.
\end{equation}
This means that the canonical commutation relation (\ref{FCR-L's}) of the OAM does not hold for the photon. Indeed, van Enk and Nienhuis \cite{Enk} once showed in a second quantized theory that the spin $\hat{\mathbf S}$ and OAM $\hat{\mathbf L}$ of the photon satisfy commutation relations,
\begin{subequations}
\begin{align}
    [\hat{S}_i, \hat{S}_j] & = 0, \label{CR-S's}\\
    [\hat{L}_i, \hat{L}_j] & = i \hbar \epsilon_{ijk} (\hat{L}_k -\hat{S}_k), \label{CR-L's}
\end{align}
\end{subequations}
respectively. These relations show that neither the spin nor the OAM can generate rotations \cite{Enk}. The conceptual separation of the spin from the OAM does not contradict \cite{Li09-1, Bliokh} the transversality condition. The usual argument \cite{Akhiezer, Simmons, Bere, Cohen, Barnett10} against such a separation is incorrect.
In fact, distinct effects of the spin and OAM were experimentally observed \cite{O'Neil, Garces} in their interactions with tiny birefringent particles trapped off axis in optical tweezers. The spin angular momentum makes the particle rotate about its own axis and the OAM makes the particle rotate about the axis of the optical beam. The conversion from spin to OAM was also observed in anisotropic \cite{Marrucci}, isotropic \cite{Zhao}, and nonlinear \cite{Mosca} media.

Commutation relation (\ref{CR-S's}) reveals that the spin of the photon lies exactly in the propagation direction \cite{Enk, Mandel}, the same as that of any other massless particles \cite{Jauch}. But it is still unclear why the commutator (\ref{CR-L's}) of the OAM depends on the spin. Particularly, what can we learn from the noncommutativity of the position? The purpose of this paper is to address these issues. The main idea is to explore the role that the constraint of transversality condition plays from a transformation point of view.

It is shown that a notion called the intrinsic system of coordinates follows from the constraint of transversality condition. What is important is that the position in the intrinsic system restores the canonical commutation relations (\ref{FCR}) with the momentum. The canonical quantum numbers that they determine are therefore in association only with the intrinsic system. The intrinsic system per se, denoted by the operator of its origin in the laboratory system of coordinates, is dependent on the helicity, the intrinsic degree of freedom. This is the reason why it is called the ``intrinsic system''. The noncommutativity (\ref{noncommute}) of the position in the laboratory system reveals the dependence of the intrinsic system on the momentum; and the spin-dependent commutator (\ref{CR-L's}) of the OAM reveals the dependence of the intrinsic system on the helicity.

The feasibility to introduce the intrinsic system lies in a quasi unitary matrix \cite{Li2008, Li09-1} that is buried in the transversality condition. It is this matrix that transforms the laboratory representation into the intrinsic representation. The wavefunction in the intrinsic representation is free of any constraints. The constraint of transversality condition in this case is transferred to the operators. Particularly, the operator of the origin of the intrinsic system is expressed solely in terms of the quasi unitary matrix and is transverse in the sense that it is perpendicular to the wavevector.
However, the transversality condition per se is not able to fully determine the quasi unitary matrix and thus the intrinsic system. In order to do so, complementary degrees of freedom \cite{Li2008, Li09-1} are necessary. Because the canonical quantum numbers are associated with the intrinsic system,
the new degrees of freedom that are expressible as a unit vector exhibit observable quantum effects.

\section{From transversality condition to quasi unitary transformation}

As is known \cite{Akhiezer, Cohen}, the momentum-space ($\mathbf k$-space) operators of the spin and the OAM about the origin of the laboratory system take the forms
\begin{subequations}\label{AM-operators-LR}
\begin{align}
  \hat{\mathbf S} & = \hbar \hat{\boldsymbol \Sigma}, \label{spin-operator-LR} \\
  \hat{\mathbf L} & =-\hat{\mathbf P} \times \hat{\mathbf X}, \label{OAM-operator-LR}
\end{align}
\end{subequations}
respectively, where $(\hat{\Sigma}_k)_{ij} =-i \epsilon_{ijk}$ with $\epsilon_{ijk}$ the
Levi-Civit\'{a} pseudotensor,
$\hat{\mathbf P} =\hbar \mathbf{k}$, $\mathbf k$ is the wavevector,
$\hat{\mathbf X}= i \nabla$
is the operator of position in the laboratory system, and $\nabla$ is the gradient operator with respect to $\mathbf k$. The $\mathbf k$-space vector wavefunction
$\mathbf{f} (\mathbf{k},t)$
on which they act satisfies the Schr\"{o}dinger equation
\begin{equation}\label{Schrodinger-eq-L}
    i \frac{\partial \mathbf{f}}{\partial t}= \omega \mathbf{f}
\end{equation}
and is constrained by the transversality condition
\begin{equation}\label{TC}
    \mathbf{w}^{T} \mathbf{f}=0,
\end{equation}
where the angular frequency $\omega =ck$ plays the role of Hamiltonian,
$k= |\mathbf{k}|$,
$\mathbf{w}= \mathbf{k}/k$,
the superscript $T$ stands for the transpose, and the convention of matrix multiplication is used for the scalar product of two vectors.
Schr\"{o}dinger equation (\ref{Schrodinger-eq-L}) together with transversality condition (\ref{TC}) is strictly equivalent to the free-space Maxwell's equations \cite{Akhiezer, Cohen}. That is to say,
if the intensities of the electric and magnetic fields that solve the free-space Maxwell's equations are written as
$\frac{1}{\sqrt{2}}(\mathbf{E}+\mathbf{E}^\ast)$
and
$\frac{1}{\sqrt{2}}(\mathbf{H}+\mathbf{H}^\ast)$,
respectively, then they are uniquely determined by the vector wavefunction that satisfies Eqs. (\ref{Schrodinger-eq-L}) and (\ref{TC}) as
\begin{subequations}\label{E-and-H}
\begin{align}
  \mathbf{E} (\mathbf{X},t) & = \frac{1}{(2 \pi)^{3/2}}
                                \int \sqrt{\frac{\hbar \omega}{\varepsilon_0}} \mathbf{f}
                                     \exp(i \mathbf{k} \cdot \mathbf{X}) d^3 k, \label{E-vec}\\
  \mathbf{H} (\mathbf{X},t) & = \frac{1}{(2 \pi)^{3/2}}
                               \int \sqrt{\frac{\hbar \omega}{\mu_0}} \mathbf{w} \times \mathbf{f}
                                    \exp(i \mathbf{k} \cdot \mathbf{X}) d^3 k.
\end{align}
\end{subequations}
Previously, it was shown \cite{Li2008, Li09-1} that the transversality condition (\ref{TC}) allows one to convert the vector wavefunction $\mathbf f$ into a two-component wavefunction $\tilde f$ by a matrix $\varpi$ as follows,
\begin{equation}\label{QUT-1}
    \tilde{f} (\mathbf{k}, t)
        =\left(
           \begin{array}{c}
             \mathbf{u}^{T} \mathbf{f} \\
             \mathbf{v}^{T} \mathbf{f} \\
           \end{array}
         \right)
        \equiv \varpi \mathbf{f} (\mathbf{k}, t),
\end{equation}
where the 2-by-3 matrix
$\varpi=\left(
          \begin{array}{c}
            \mathbf{u}^{T} \\
            \mathbf{v}^{T} \\
          \end{array}
        \right)
$
consists of two time-independent row vectors $\mathbf{u}^{T}$ and $\mathbf{v}^{T}$
that are perpendicular to each other and form a local righthand Cartesian system with $\mathbf w$,
\begin{subequations}\label{triad}
\begin{align}
  \mathbf{v}^T \mathbf{u}
      =\mathbf{w}^T \mathbf{v}
      =\mathbf{u}^T \mathbf{w}                    & =0, \\
  \mathbf{u}^T \mathbf{u}=\mathbf{v}^T \mathbf{v} & =1, \\
  \mathbf{u} \times \mathbf{v}                    & = \mathbf{w}. \label{righthand}
\end{align}
\end{subequations}
The two-component wavefunction $\tilde f$ is free of any constraints such as Eq. (\ref{TC}). Multiplying both sides of Eq. (\ref{Schrodinger-eq-L}) by $\varpi$ from the left and making use of Eq. (\ref{QUT-1}), one has the following Schr\"{o}dinger equation of the two-component wavefunction,
\begin{equation}\label{Schrodinger-eq-I}
    i \frac{\partial \tilde{f}}{\partial t}= \omega \tilde{f}.
\end{equation}

The matrix $\varpi$ in Eq. (\ref{QUT-1}) performs a quasi unitary transformation. Firstly, it is not difficult to show that
$\varpi^T \varpi =I_3-\mathbf{w} \mathbf{w}^T$,
where
$\varpi^T= \left(
             \begin{array}{cc}
               \mathbf{u} & \mathbf{v} \\
             \end{array}
           \right)
$
is a 3-by-2 matrix and $I_3$ denotes the 3-by-3 unit matrix. With the help of Eq. (\ref{TC}), one has
\begin{equation*}
    \varpi^T \varpi \mathbf{f}=\mathbf{f}.
\end{equation*}
Remembering that $\varpi$ and therefore $\varpi^T \varpi$ always acts on the vector wavefunction $\mathbf f$ that is constrained by the transversality condition (\ref{TC}), one may rewrite the above equation simply as
\begin{equation}\label{unitarity-1}
    \varpi^T \varpi =I_3.
\end{equation}
Multiplying both sides of Eq. (\ref{QUT-1}) by $\varpi^T$ from the left and considering this relation, one gets
\begin{equation}\label{QUT-2}
    \mathbf{f} =\varpi^T \tilde{f}.
\end{equation}
It says that the matrix $\varpi^T$ transforms a two-component wavefunction into a vector wavefunction. Secondly, it is easy to prove that
\begin{equation}\label{unitarity-2}
    \varpi \varpi^T =I_2,
\end{equation}
where $I_2$ denotes the 2-by-2 unit matrix. Eqs. (\ref{unitarity-1}) and (\ref{unitarity-2}) express the quasi unitarity \cite{Golub} of the transformation matrix $\varpi$. They guarantee that the norm of the wavefunction remains unchanged under the transformation,
$\tilde{f}^{\dag} \tilde{f} =\mathbf{f}^{\dag} \mathbf{f}$.
$\varpi^T$ is the Moore-Penrose pseudo inverse of $\varpi$, and vice versa.

\section{Intrinsic and laboratory representations}\label{representations}

Along with the transformation of the vector wavefunction into the two-component wavefunction via Eq. (\ref{QUT-1}), the spin operator (\ref{spin-operator-LR}) that acts on the vector wavefunction is transformed into \cite{Sakurai}
\begin{equation}\label{defining-spin-IR}
    \hat{\mathbf s}
   =\varpi \hat{\mathbf S} \varpi^{T}
   =\hbar \varpi \hat{\boldsymbol \Sigma} \varpi^{T},
\end{equation}
which acts on the two-component wavefunction. Upon decomposing the vector operator $\hat{\boldsymbol \Sigma}$ in the local Cartesian system $uvw$ as
\begin{equation*}
    \hat{\boldsymbol \Sigma}
       =(\mathbf{u}^T \hat{\boldsymbol \Sigma}) \mathbf{u}
       +(\mathbf{v}^T \hat{\boldsymbol \Sigma}) \mathbf{v}
       +(\mathbf{w}^T \hat{\boldsymbol \Sigma}) \mathbf{w}
\end{equation*}
and making use of Eqs. (\ref{triad}), one gets
\begin{equation}\label{spin-operator-IR}
    \hat{\mathbf s}= \hbar \hat{\sigma}_3 \mathbf{w},
\end{equation}
where the helicity operator
\begin{equation}\label{sigma3}
    \hat{\sigma}_3=\varpi (\mathbf{w}^T \hat{\mathbf \Sigma}) \varpi^{T}
                  =\left(
                     \begin{array}{cc}
                       0 & -i \\
                       i &  0 \\
                     \end{array}
                   \right)
\end{equation}
is one of the Pauli matrices. Eq. (\ref{spin-operator-IR}) reveals an important result that the photon spin lies entirely along the direction of the wavevector \cite{Enk, Jauch, Mandel}. The present approach to Eq. (\ref{spin-operator-IR}) demonstrates that it is this property of the spin that underlies the transversality condition. The inverse transformation of Eq. (\ref{defining-spin-IR}) gives for the spin operator acting on the vector wavefunction,
\begin{equation}\label{SO-helicity-LR}
    \hat{\mathbf S}
        =\varpi^{T} \hat{\mathbf s} \varpi
        =\hbar (\mathbf{w}^T \hat{\boldsymbol \Sigma}) \mathbf{w},
\end{equation}
where Eqs. (\ref{unitarity-1}), (\ref{unitarity-2}), (\ref{spin-operator-IR}), and (\ref{sigma3}) are used. Commutation relation (\ref{CR-S's}) of the spin follows directly from Eq. (\ref{SO-helicity-LR}). Here we arrive at it without resorting to the second quantization.

Next let us turn our attention to the OAM. Acting on the vector wavefunction of three components, the operators of the position and momentum should be understood as
\begin{equation}
  \hat{\mathbf X}= i \nabla \otimes I_3, \hspace{5pt} \hat{\mathbf P}= \hbar \mathbf{k} \otimes I_3,
\end{equation}
respectively.
Along with the transformation of the vector wavefunction into the two-component wavefunction, they are transformed into
\begin{subequations}
\begin{align}
  \hat{\mathbf x} & =\varpi \hat{\mathbf X} \varpi^{T}=\hat{\boldsymbol \xi} +\hat{\mathbf b}, \label{operator-x} \\
  \hat{\mathbf p} & =\varpi \hat{\mathbf P} \varpi^{T}=\hbar \mathbf{k} \otimes I_2, \label{operator-p}
\end{align}
\end{subequations}
respectively, where
$\hat{\boldsymbol \xi} =i \nabla \otimes I_2$
and
\begin{equation}\label{b}
    \hat{\mathbf b} =i \varpi (\nabla \varpi^{T}).
\end{equation}
The position operator now splits into two parts. The first part $\hat{\boldsymbol \xi}$ looks like $\hat{\mathbf X}$ but acts on the two-component wavefunction. Because no constraints such as Eq. (\ref{TC}) exist for the two-component wavefunction, the Cartesian coordinates of $\hat{\boldsymbol \xi}$ commute,
\begin{equation}\label{CR-xi}
    [\hat{\xi}_i, \hat{\xi}_j] =0.
\end{equation}
Noticing that the Cartesian components of the momentum $\hat{\mathbf p}$ also commute,
\begin{equation}\label{CR-p}
    [\hat{p}_i, \hat{p}_j] =0,
\end{equation}
this part is canonically conjugate to the momentum \cite{Cohen}, obeying the following commutation relation,
\begin{equation}\label{CR-xi-with-p}
    [\hat{\xi}_i,\hat{p}_j] =i \hbar \delta_{ij}.
\end{equation}
The fundamental quantum conditions (\ref{CR-xi})-(\ref{CR-xi-with-p}) are very important. They indicate that the two-component wavefunction is defined in such a system of coordinates in which the position is represented by operator $\hat{\boldsymbol \xi}$. In other words, the two-component wavefunction is defined in a system in which the fundamental quantum conditions restore. This is to be compared with the vector wavefunction that is defined in the laboratory system $\mathbf X$ in which the fundamental quantum conditions break down.

That the operator $\hat{\boldsymbol \xi}$ takes on the same gradient form as the operator $\hat{\mathbf X}$ does indicates that the axes of system $\boldsymbol \xi$ are parallel to those of the laboratory system. So the physical meaning of system $\boldsymbol \xi$ is expressed only by operator $\hat{\mathbf b}$, the second part of $\hat{\mathbf x}$. As can be seen from Eq. (\ref{operator-x}), this part represents the origin of the system $\boldsymbol \xi$ in the laboratory system.
It is totally determined by $\varpi$. It is Hermitian,
\begin{equation*}
    \hat{\mathbf b}^{\dag} =-i (\nabla \varpi) \varpi^{T}
                                 = i \varpi (\nabla \varpi^{T})
                                 = \hat{\mathbf b},
\end{equation*}
by virtue of Eq. (\ref{unitarity-2}). Moreover, being commutative with the Hamiltonian, it is a constant of motion. Its Cartesian components commute,
\begin{equation}\label{CR-Xi}
    [\hat{b}_i, \hat{b}_j]=0.
\end{equation}
But the problem is that the requirements (\ref{triad}) cannot fully determine $\varpi$ up to a rotation about the wavevector \cite{Mandel}. In other words, the constraint of transversality condition per se is not able to fully determine the system $\boldsymbol \xi$. In order to do so, one needs additional degrees of freedom.
It was once shown \cite{Li2008, Li09-1} that a constant unit vector can meet the need. Indeed, denoting by $\mathbf I$ any constant unit vector, it is easy to check that the following two unit vectors do satisfy Eqs. (\ref{triad}),
\begin{equation}\label{QUT basis}
    \mathbf{u}_{\mathbf{I}} =\mathbf{v}_{\mathbf{I}} \times \frac{\mathbf k}{k},
         \hspace{5pt}
    \mathbf{v}_{\mathbf{I}} =\frac{\mathbf{I} \times \mathbf{k}}
         {|\mathbf{I} \times \mathbf{k}|}.
\end{equation}
To reflect this $\mathbf I$-dependence, we rewrite the transformation matrix explicitly as
\begin{equation}\label{QUM}
    \varpi_{\mathbf I}=\left(
                         \begin{array}{c}
                           \mathbf{u}^{T}_{\mathbf I} \\
                           \mathbf{v}^{T}_{\mathbf I} \\
                         \end{array}
                       \right).
\end{equation}
In this case, Eqs. (\ref{unitarity-1}) and (\ref{unitarity-2}) for its quasi unitarity take the forms
\begin{subequations}\label{unitarity}
\begin{align}
  \varpi^{T}_{\mathbf I} \varpi_{\mathbf I} & = I_3, \label{unitarity-I1} \\
  \varpi_{\mathbf I} \varpi^{T}_{\mathbf I} & = I_2, \label{unitarity-I2}
\end{align}
\end{subequations}
respectively. Accordingly, Eq. (\ref{QUT-1}) for the two-component wavefunction takes the form
\begin{equation}\label{tilde{f}-I}
    \tilde{f}_{\mathbf I}= \varpi_{\mathbf I} \mathbf{f}.
\end{equation}
Substituting Eq. (\ref{QUM}) into Eq. (\ref{b}) and denoting the resultant by $\hat{\mathbf b}_{\mathbf I}$, one gets
\begin{equation}\label{b-I}
    \hat{\mathbf b}_{\mathbf I}
        =\frac{\mathbf{I} \cdot \mathbf{k}}{k |\mathbf{I} \times \mathbf{k}|} \mathbf{v}_{\mathbf I}
         \otimes \hat{\sigma}_3.
\end{equation}
The one-to-one correspondence between Eqs. (\ref{tilde{f}-I}) and (\ref{b-I}) shows that the degree of freedom $\mathbf I$ plays the role of specifying the system $\boldsymbol \xi$ in which the two-component wavefunction is defined.

It is seen from Eq. (\ref{b-I}) that the origin of the system $\boldsymbol \xi$ in the laboratory system is dependent on the helicity $\hat{\sigma}_3$, the intrinsic degree of freedom. Due to this peculiar property, we will refer to the system $\boldsymbol \xi$ as the intrinsic system in order to distinguish it from the laboratory system. Considering that the two-component wavefunction is defined in the intrinsic system, we will refer to the two-component representation as the intrinsic representation. Correspondingly, the vector representation will be referred to as the laboratory representation. Evidently, the degree of freedom to specify the intrinsic system amounts to the degree of freedom to specify the intrinsic representation. In so specified intrinsic representation, Eq. (\ref{operator-x}) for the operator of position in the laboratory system takes the form
\begin{equation}\label{x-I}
    \hat{\mathbf x}_{\mathbf I}=\hat{\boldsymbol \xi} +\hat{\mathbf b}_{\mathbf I}.
\end{equation}
With the help of Eqs. (\ref{CR-xi}) and (\ref{CR-Xi}), it is easy to find
\begin{equation}\label{CR-position}
    \hat{\mathbf x}_{\mathbf I} \times \hat{\mathbf x}_{\mathbf I}
        =i \nabla \times \hat{\mathbf b}_{\mathbf I},
\end{equation}
which says that the noncommutativity of the position in the laboratory system \cite{Pryce, Skagerstam, Berard2006} originates in the dependence of the intrinsic system on the momentum.

Different from the laboratory wavefunction, the intrinsic wavefunction is not unique for a particular radiation field. Here the degree of freedom to determine the intrinsic wavefunction via Eq. (\ref{tilde{f}-I}) is analogous to the gauge degree of freedom to determine the gauge potentials in classical theory \cite{Stratton} as we will see below.

\section{Quantum analog to the gauge degree of freedom}

To this end, let us see how the change of $\mathbf I$ affects the intrinsic wavefunction.
When $\mathbf I$ is changed into a different value, say $\mathbf{I}'$, the intrinsic wavefunction for the same radiation field becomes
\begin{equation}\label{tilde{f}-I'}
    \tilde{f}_{\mathbf{I}'}= \varpi_{\mathbf{I}'} \mathbf{f},
\end{equation}
where
$
\varpi_{\mathbf{I}'}=\left(
                       \begin{array}{c}
                         \mathbf{u}^{T}_{\mathbf{I}'} \\
                         \mathbf{v}^{T}_{\mathbf{I}'} \\
                       \end{array}
                     \right)
$
and
\begin{equation*}
    \mathbf{u}_{\mathbf{I}'} =\mathbf{v}_{\mathbf{I}'} \times \frac{\mathbf k}{k},
         \hspace{5pt}
    \mathbf{v}_{\mathbf{I}'} =\frac{\mathbf{I}' \times \mathbf{k}}
         {|\mathbf{I}' \times \mathbf{k}|}.
\end{equation*}
As remarked earlier, the orthonormal vectors
$\mathbf{u}_{\mathbf{I}'}$ and $\mathbf{v}_{\mathbf{I}'}$
that make up the new transformation matrix $\varpi_{\mathbf{I}'}$ are related to the old orthonormal vectors $\mathbf{u}_{\mathbf I}$ and $\mathbf{v}_{\mathbf I}$ by a rotation about $\mathbf w$. Such a rotation can be expressed as follows,
\begin{subequations}\label{rotation-SO(3)}
\begin{align}
  \mathbf{u}_{\mathbf{I}'} & = \mathbf{u}_{\mathbf I} \cos \phi +\mathbf{v}_{\mathbf I} \sin \phi, \\
  \mathbf{v}_{\mathbf{I}'} & =-\mathbf{u}_{\mathbf I} \sin \phi +\mathbf{v}_{\mathbf I} \cos \phi,
\end{align}
\end{subequations}
where $\phi$ denotes the relevant rotation angle.
These two equations can be integrated in terms of the transformation matrices $\varpi_{\mathbf{I}'}$ and $\varpi_{\mathbf I}$ into
\begin{equation}\label{rotation-Pi}
    \varpi_{\mathbf{I}'}=D \varpi_{\mathbf I},
\end{equation}
where
$
    D=\left(
        \begin{array}{cc}
           \cos\phi & \sin\phi \\
          -\sin\phi & \cos\phi \\
        \end{array}
      \right)
$ is the rotation matrix, which can be expressed in terms of the helicity operator $\hat{\sigma}_3$ as
\begin{equation}\label{RO}
    D =\exp \left(i \hat{\sigma}_3 \phi \right).
\end{equation}
Substituting Eq. (\ref{rotation-Pi}) into Eq. (\ref{tilde{f}-I'}) and making use of Eqs. (\ref{tilde{f}-I}) and (\ref{RO}), we have
\begin{equation}\label{OT}
    \tilde{f}_{\mathbf{I}'}
   =\exp(i \hat{\sigma}_3 \phi)\tilde{f}_{\mathbf I}.
\end{equation}
The rotation angle $\phi$ is determined as follows.

The barycenter of the radiation field in the laboratory system should be invariant under the change of $\mathbf I$. This means that
\begin{equation}\label{sb}
    \tilde{f}^{\dag}_{\mathbf{I}'} (i \nabla+ \hat{\mathbf b}_{\mathbf{I}'}) \tilde{f}_{\mathbf{I}'}
   =\tilde{f}^{\dag}_{\mathbf{I}}  (i \nabla+ \hat{\mathbf b}_{\mathbf{I}} ) \tilde{f}_{\mathbf{I}},
\end{equation}
by virtue of Eq. (\ref{x-I}), where
$\hat{\mathbf b}_{\mathbf{I}'}= i \varpi_{\mathbf{I}'} (\nabla \varpi^{T}_{\mathbf{I}'})$.
Upon substituting Eq. (\ref{OT}) into Eq. (\ref{sb}), we find
\begin{equation*}
    \tilde{f}^{\dag}_{\mathbf{I}} (\hat{\mathbf b}_{\mathbf{I}'}- \hat{\mathbf b}_{\mathbf{I}}) \tilde{f}_{\mathbf{I}}
   =\tilde{f}^{\dag}_{\mathbf{I}} [(\nabla \phi) \otimes \hat{\sigma}_3] \tilde{f}_{\mathbf{I}},
\end{equation*}
which is equivalent to
\begin{equation}\label{GT}
    \hat{\mathbf b}_{\mathbf{I}'}- \hat{\mathbf b}_{\mathbf{I}}
   =(\nabla \phi) \otimes \hat{\sigma}_3,
\end{equation}
from the arbitrariness of the laboratory wavefunction $\mathbf f$ and of the unit vector $\mathbf I$. Obviously, the rotation angle $\phi$ satisfying this equation is dependent on the wavevector.

It is well known in classical electromagnetic theory \cite{Stratton} that the electric and magnetic fields of any radiation field in free space can be expressed in terms of four gauge potentials. In that expression, there exists a gauge degree of freedom in the sense that the electric and magnetic fields are invariant under a gauge transformation of the potentials. Because only two of the four gauge potentials are truly independent \cite{Cohen}, the intrinsic wavefunction given by Eq. (\ref{tilde{f}-I}) can be regarded as the quantum analog to the gauge potentials; the degree of freedom $\mathbf I$ as the quantum analog to the gauge degree of freedom; Eq. (\ref{OT}) as the quantum analog to the gauge transformation of the potentials, where the ``gauge function'' $\phi$ is determined by Eq. (\ref{GT}).

\section{Canonical quantization of radiation fields}\label{CQ}

It is observed that in the intrinsic representation the helicity operator $\hat{\sigma}_3$ is independent of the canonical variables $\hat{\boldsymbol \xi}$ and $\hat{\mathbf p}$. This fact allows us to canonically quantize a radiation field in the intrinsic representation.
On one hand, commuting with the Hamiltonian, the helicity is a constant of motion. It has eigenvalues
$\sigma_3= \pm 1$,
corresponding to normalized eigenfunctions
\begin{equation}\label{eigen spinors}
    \tilde{\alpha}_{+1}=\frac{1}{\sqrt{2}} \left(
                                             \begin{array}{c}
                                               1 \\
                                               i \\
                                             \end{array}
                                           \right),
\hspace{10pt}
    \tilde{\alpha}_{-1}=\frac{1}{\sqrt{2}} \left(
                                              \begin{array}{c}
                                                i \\
                                                1 \\
                                              \end{array}
                                           \right),
\end{equation}
respectively.
On the other hand, a set of three commuting canonical variables, also being constants of motion, can be deduced from the canonical commutation relations (\ref{CR-xi})-(\ref{CR-xi-with-p}). So we have a complete set of four quantum numbers to characterize the complete set of eigenfunctions in the intrinsic representation. One is the helicity quantum number $\sigma_3$. The other three, denoted collectively by $q$ and called the canonical quantum numbers, depend on the choice of the set of three commuting canonical variables. Letting be $f_q$ the simultaneous normalized eigenfunctions of the commuting canonical variables, we may write the complete orthonormal set of eigenfunctions in the intrinsic representation as
$$
\tilde{f}_{\sigma_3 q} (\mathbf{k}, t)= \tilde{\alpha}_{\sigma_3} f_{q} (\mathbf{k}) \exp(-i \omega t)
$$
and their orthonormality relation as
\begin{equation}\label{orthonormal-ftilde}
    \int \tilde{f}^{\dag}_{\sigma'_3 q'} \tilde{f}_{\sigma_3 q} d^3 k
   =\delta_{\sigma'_{3} \sigma_{3}} \delta_{q' q},
\end{equation}
where the Kronecker $\delta_{q' q}$ should be replaced with the Dirac $\delta$-function for continuous canonical quantum numbers. The following are three commonly used sets of canonical quantum numbers.

First of all, the momentum is a constant of motion. According to commutation relation (\ref{CR-p}), one can choose the three Cartesian components of the momentum as the commuting canonical variables. Denoting by $\hbar \mathbf{k}_0$ the eigen momentum, one has the following $q$-dependent factors for the complete set of eigenfunctions,
\begin{equation}\label{pw-IR}
    f_{q} (\mathbf{k}) =\delta^3 (\mathbf{k}-\mathbf{k}_0),
\end{equation}
where $q=\mathbf{k}_0$. Their orthonormality relation reads
\begin{equation}\label{orthonormal-p}
    \int f^{\ast}_{q'} f_{q} d^3 k
   =\delta^3 (\mathbf{k}'_0-\mathbf{k}_0),
\end{equation}
where $q'=\mathbf{k}'_0$.

Secondly, now that the position in the intrinsic system is canonically conjugate to the momentum, the OAM about the origin of the intrinsic system,
$\hat{\boldsymbol \lambda}=-\hat{\mathbf p} \times \hat{\boldsymbol \xi}$,
obeys the canonical commutation relation (\ref{FCR-L's}). That is to say, one has
\begin{equation}\label{CR-lambda}
    [\hat{\lambda}_i, \hat{\lambda}_j] =i \hbar \epsilon_{ijk} \hat{\lambda}_k.
\end{equation}
Since this OAM commutes with the Hamiltonian,
\begin{equation}\label{CR-lambda and omega}
    [\hat{\boldsymbol \lambda}, \omega] =0,
\end{equation}
one can also choose
$\omega$, $\hat{\boldsymbol \lambda}^2$, and $\hat{\lambda}_3$
as the commuting canonical variables.
It is well known \cite{Cohen} that the simultaneous normalized eigenfunctions of $\hat{\boldsymbol \lambda}^2$ and $\hat{\lambda}_3$ in $\mathbf k$-space are the spherical surface harmonics ,
\begin{equation*}
    Y_{\lambda \mu} (\mathbf{w})
   =\left\{ \frac{2\lambda +1}{4 \pi} \frac{(\lambda-\mu)!}{(\lambda+\mu)!} \right\}^{1/2}
    P_{\lambda}^{\mu} (\cos \vartheta) e^{i \mu \varphi},
\end{equation*}
which satisfy the following eigenvalue equations,
\begin{subequations}
\begin{align}
  \hat{\boldsymbol \lambda}^2 Y_{\lambda \mu}= \lambda(\lambda+1) \hbar^2 Y_{\lambda \mu}, &
               \hspace{5pt} \lambda=0, 1, 2... \\
  \hat{\lambda}_3 Y_{\lambda \mu}= \mu \hbar Y_{\lambda \mu},                              &
               \hspace{5pt} \mu=    0, \pm 1, \pm 2 ... \pm \lambda.
\end{align}
\end{subequations}
Their orthonormality relation assumes the form
\begin{equation*}
    \int Y_{\lambda' \mu'}^{\ast} Y_{\lambda \mu} \sin \vartheta d \vartheta d \varphi
   =\delta_{\lambda' \lambda} \delta_{\mu' \mu}.
\end{equation*}
As a result, the expected $q$-dependent factors for the complete set of eigenfunctions in this case take the form
\begin{equation}\label{spherical eigenfunction}
    f_{q} (\mathbf{k})
   =\frac{\delta(k-k_0)}{\sqrt{k_0 \omega_0}} Y_{\lambda \mu} (\mathbf{w}),
\end{equation}
where $q=\{ \omega_0, \lambda, \mu \}$, $\omega_0$ is the eigen energy, and $k_0=\omega_0/c$. They have the following orthonormality relation,
\begin{equation}\label{orthonormal-h}
    \int f^{\ast}_{q'} f_{q} d^3 k
   =\delta(\omega'_0-\omega_0) \delta_{\lambda' \lambda} \delta_{\mu' \mu},
\end{equation}
where $q'=\{\omega'_0, \lambda', \mu' \}$.

At last, noticing that $\hat{p}_3$ and $\hat{\lambda}_3$ commute, a third choice of the commuting canonical variables can be
$\omega$, $\hat{p}_3$, and $\hat{\lambda}_3$ as well.
$\hat{p}_3$ and $\hat{\lambda}_3$ have the following simultaneous normalized eigenfunctions in circular cylindrical coordinates,
\begin{equation*}
    X_{k_{30} \mu} (\varphi, k_3)=\frac{1}{\sqrt{2 \pi}} \delta (k_3-k_{30}) e^{i \mu \varphi},
    \hspace{5pt} \mu=0, \pm1, \pm2...
\end{equation*}
with eigenvalues $k_{30} \hbar$ and $\mu \hbar$, respectively. The $q$-dependent factors for the corresponding complete set of eigenfunctions are given by
\begin{equation}\label{CF}
    f_{q} (\mathbf{k})
        =\frac{\sqrt{\omega_0}}{c k_{\rho 0}} \delta(k_{\rho}-k_{\rho 0}) X_{k_{30} \mu} (\varphi, k_3),
\end{equation}
which describe diffraction-free light beams in position space \cite{Wang}, where $q= \{\omega_0, k_{30}, \mu)$ and $k_{\rho 0}= (k^2_0-k^2_{30})^{1/2}$. They satisfy the following orthonormality relation,
\begin{equation}\label{orthonormal-d}
    \int f^{\ast}_{q'} f_{q} d^3 k
   =\delta(\omega'_0-\omega_0) \delta (k'_{30}-k_{30}) \delta_{\mu' \mu},
\end{equation}
where $q'=\{\omega'_0, k'_{30}, \mu' \}$.

The complete set of eigenfunctions in the intrinsic representation has nothing to do with the degree of freedom $\mathbf I$. Nevertheless, the intrinsic wavefunction for a particular radiation field is dependent on $\mathbf I$ as is shown by Eq. (\ref{tilde{f}-I}). When one expands this wavefunction in a complete set, the expansion coefficient must depend on $\mathbf I$,
\begin{equation}\label{expansion-tilde-f}
    \tilde{f}_{\mathbf I}=\sum_{\sigma_3, q} a_{\mathbf{I} \sigma_3 q} \tilde{f}_{\sigma_3 q},
\end{equation}
where it is assumed that the summation over $q$ includes the integration over continuous canonical quantum numbers. To see what is meant by this $\mathbf I$-dependence, we convert Eq. (\ref{tilde{f}-I}) into
$\mathbf{f}= \varpi^{T}_{\mathbf I} \tilde{f}_{\mathbf I}$
with the help of Eq. (\ref{unitarity-I1}) and substitute Eq. (\ref{expansion-tilde-f}) to get
\begin{equation}\label{expansion}
    \mathbf{f}=\sum_{\sigma_3, q} a_{\mathbf{I} \sigma_3 q} \mathbf{f} _{\mathbf{I} \sigma_3 q},
\end{equation}
where
\begin{equation}\label{CS-LR}
\mathbf{f}_{\mathbf{I} \sigma_3 q} =\varpi^{T}_{\mathbf{I}} \tilde{f}_{\sigma_3 q}.
\end{equation}
Eq. (\ref{expansion}) demonstrates that when the quantum numbers $\sigma_3$ and $q$ run over all their possible values, any constant unit vector $\mathbf I$ determines, through Eq. (\ref{CS-LR}), a complete set of eigenfunctions to span the laboratory representation. That is to way, given a set of four quantum numbers $\sigma_3$ and $q$, we still have a degree of freedom to choose the complete set of eigenfunctions in the laboratory representation. Inserting Eq. (\ref{unitarity-I2}) into the left side of Eq. (\ref{orthonormal-ftilde}) and making use of Eq. (\ref{CS-LR}), one has the following orthonormality relation,
\begin{equation}\label{orthonormal-fvector}
    \int \mathbf{f}^{\dag}_{\mathbf{I} \sigma'_3 q'} \mathbf{f}_{\mathbf{I} \sigma_3 q} d^3 k
   =\delta_{\sigma'_{3} \sigma_{3}} \delta_{q' q}.
\end{equation}
The expansion coefficient is thus obtained from Eq. (\ref{expansion}) to be
\begin{equation}\label{coefficients}
    a_{\mathbf{I} \sigma_3 q}
   =\int \mathbf{f}^{\dag}_{\mathbf{I} \sigma_3 q} \mathbf{f} d^3 k.
\end{equation}
Because a laboratory wavefunction uniquely describes a radiation field via Eqs. (\ref{E-and-H}), Eq. (\ref{CS-LR}) reveals an important fact that the eigen excitation of the radiation field that it defines is determined not only by a set of four quantum numbers but also by the degree of freedom $\mathbf I$. The $\mathbf I$-dependence of the expansion coefficient (\ref{coefficients}) just reflects the dependence of the eigen excitation on this degree of freedom.
After second quantization, the expansion coefficient becomes the annihilation operator. In a word, Eq. (\ref{expansion-tilde-f}) or (\ref{expansion}) expresses the canonical quantization of the radiation field.

The dependence of the eigen excitation on the degree of freedom $\mathbf I$ can be understood as follows. The canonical commutation relations (\ref{CR-xi})-(\ref{CR-xi-with-p}) in the intrinsic system mean that the canonical quantum numbers $q$ they determine are associated only with the intrinsic system. Moreover, the helicity quantum number $\sigma_3$ is just an intrinsic degree of freedom. In order to completely determine an eigen excitation, the intrinsic system with which the canonical quantum numbers are associated must be unambiguously specified in the laboratory system. This is done by the degree of freedom $\mathbf I$. In fact, it is not difficult to show that in any eigen excitation that has the intrinsic wavefunction $\tilde{f}_{\sigma_3 q}$ discussed before, the expectation value of operator $\hat{\boldsymbol \xi}$ vanishes,
\begin{equation}\label{<xi>}
    \langle \hat{\boldsymbol \xi} \rangle_{\sigma_3 q}
  \equiv \frac{\int \tilde{f}^{\dag}_{\sigma_3 q} \hat{\boldsymbol \xi} \tilde{f}_{\sigma_3 q} d^3 k}
         {\int \tilde{f}^{\dag}_{\sigma_3 q} \tilde{f}_{\sigma_3 q} d^3 k}=0.
\end{equation}
This means that for the eigen excitation the intrinsic system reduces to its barycenter system \cite{Yang} and the $\mathbf I$-dependent operator $\hat{\mathbf b}_{\mathbf I}$ represents its barycenter in the laboratory system. This is why the degree of freedom $\mathbf I$ can be used to explain \cite{Li09-2} the spin Hall effect of light \cite{Hosten}. Let us see how the change of $\mathbf I$ affects the eigen excitation of the radiation field.

We know that the eigenfunction $\tilde{f}_{\sigma_3 q}$ in the intrinsic representation satisfies
\begin{equation}\label{EH-IR}
    \hat{\sigma}_3 \tilde{f}_{\sigma_3 q} =\sigma_3 \tilde{f}_{\sigma_3 q}.
\end{equation}
For a particular unit vector $\mathbf I$, the eigenfunction in the laboratory representation is given by Eq. (\ref{CS-LR}).
When $\mathbf I$ is changed into a different value, say $\mathbf{I}'$, the corresponding eigenfunction in the laboratory representation is given by
\begin{equation}\label{CS-LR'}
    \mathbf{f}_{\mathbf{I}' \sigma_3 q} =\varpi^{T}_{\mathbf{I}'} \tilde{f}_{\sigma_3 q}.
\end{equation}
Substituting Eqs. (\ref{rotation-Pi}) and (\ref{RO}) into Eq. (\ref{CS-LR'}) and noticing the eigenvalue equation (\ref{EH-IR}), we get
\begin{equation}\label{CS-MR'-MR}
    \mathbf{f}_{\mathbf{I}' \sigma_3 q} =\exp(-i \sigma_3 \phi) \mathbf{f}_{\mathbf{I} \sigma_3 q}.
\end{equation}
It shows that when the unit vector $\mathbf I$ is changed with the quantum numbers $\sigma_3$ and $q$ remaining unchanged, the eigenfunction in the laboratory representation will acquire a phase. Depending on the helicity $\sigma_3$ as well as the wavevector through the rotation angle $\phi$, this phase will substantially affect the electric and magnetic fields of the eigen excitation as can be seen from Eqs. (\ref{E-and-H}).

\section{Quantum theory of the angular momentum}

In Section \ref{representations} we derived the commutation relation (\ref{CR-S's}) of the spin angular momentum. Now we are in a position to derive the commutation relation (\ref{CR-L's}) of the OAM in the intrinsic representation.
The same as the operator (\ref{x-I}) of the position in the laboratory system, the operator of the OAM about the origin of the laboratory system also splits into two parts,
\begin{equation}\label{OAM operator-IR}
    \hat{\mathbf l}=\varpi^{\dag}_{\mathbf I} \hat{\mathbf L} \varpi_{\mathbf I}
                   =\hat{\boldsymbol \lambda} +\hat{\mathbf m}.
\end{equation}
The first part $\hat{\boldsymbol \lambda}$ is the OAM of the photon about the origin of the intrinsic system. The second part
\begin{equation}\label{OAM-B-I}
    \hat{\mathbf m}
   \equiv \hbar \hat{\mathbf b}_{\mathbf I} \times \mathbf{k}
   =\hbar \frac{\mathbf{I} \cdot \mathbf{k}}{|\mathbf{I} \times \mathbf{k}|}
    \mathbf{u}_{\mathbf I} \otimes \hat{\sigma}_3
\end{equation}
is the OAM of the photon concentrated at the origin of the intrinsic system about the origin of the laboratory system. Obviously, $\hat{\mathbf m}$ is dependent on the helicity.
This not only explains why the entire OAM depends on the helicity \cite{Li09-1}, but also helps us to understand why the total angular momentum cannot be generally separated into helicity-dependent spin and helicity-independent OAM \cite{Barnett-A}.
Like $\hat{\mathbf b}_{\mathbf I}$, $\hat{\mathbf m}$ is also a constant of motion,
\begin{equation}\label{CR-Lambda and omega}
    [\hat{\mathbf m}, \omega] =0,
\end{equation}
and its Cartesian components commute,
\begin{equation}\label{CR-Lambda}
    [\hat{m}_i,  \hat{m}_j]=0.
\end{equation}
From Eqs. (\ref{CR-lambda and omega}) and (\ref{CR-Lambda and omega}) it follows that the entire OAM is a constant of motion, too. With the help of Eqs. (\ref{CR-lambda}) and (\ref{CR-Lambda}), straightforward calculations yield
\begin{equation}\label{CR-OAM}
    [\hat{l}_i, \hat{l}_j] =i \hbar \epsilon_{ijk} (\hat{l}_k-\hat{s}_k).
\end{equation}
It is the commutation relation (\ref{CR-L's}) of the OAM that was found by van Enk and Nienhuis \cite{Enk} when the inverse transformations of (\ref{OAM operator-IR}) and (\ref{defining-spin-IR}) are taken into account. Here we arrive at it without resorting to the second quantization. The present approach to Eq. (\ref{CR-OAM}) demonstrates that it is the helicity dependence of the intrinsic system that makes the commutator (\ref{CR-L's}) of the OAM depend on the spin.

\section{Concluding remarks}

In summary, we introduced from the constraint of transversality condition the notion of photon's intrinsic system in which the position is canonically conjugate to the momentum. Moreover we extracted from the same constraint a complementary degree of freedom that together with the helicity and the momentum determines the intrinsic system completely. The newly identified degree of freedom per se plays the role of specifying the intrinsic representation. Since the canonical quantum numbers that are determined by the fundamental quantum conditions (\ref{CR-xi})-(\ref{CR-xi-with-p}) are associated with the intrinsic system, the new degree of freedom has observable quantum effects \cite{Yang} and is responsible for the spin Hall effect \cite{Li09-2, Hosten}. Due to the dependence of the intrinsic system on the momentum, the Cartesian components of the position in the laboratory system do not commute. This noncommutativity in the laboratory representation is hidden behind the transversality condition on the laboratory wavefunction.

It is also noted that the OAM of the photon about the origin of the laboratory system is the OAM of the photon concentrated at the origin of the intrinsic system plus the OAM of the photon about the origin of the intrinsic system. According to Goldstein \cite{Goldstein}, the intrinsic system can be regarded simply as the barycenter system. This is in agreement with the observation that the intrinsic system of the eigen excitation reduces to its barycenter system. What deserves emphasizing is that this barycenter is perpendicular to the momentum due to the constraint of transversality condition.

\section*{Acknowledgments}

The author is indebted to Vladimir Fedoseyev and Zihua Xin for their helpful discussions. This work was supported in part by the National Natural Science Foundation of China (60877055).

\end{document}